\documentclass{iopart}
\usepackage{iopams}
\usepackage{amsthm}
\usepackage{epsf}
\usepackage[all]{xy}
\begin{document}
\newcommand{\Z}{\mathbb{Z}}
\newcommand{\R}{\mathbb{R}}
\title{Cohomology of matching rules}
\author{P Kalugin}
\address{Laboratoire de Physique des Solides, B\^at 510, 91405 Orsay, France}
\begin{abstract}
Quasiperiodic patterns described by polyhedral ``atomic surfaces''
and admitting matching rules are considered. It is shown that the
cohomology ring of the continuous hull of such patterns is
isomorphic to that of the complement of a torus $T^N$ to an
arrangement $A$ of thickened affine tori of codimension two.
Explicit computation of Betti numbers for several two-dimensional
tilings and for the icosahedral Ammann-Kramer tiling confirms in
most cases the results obtained previously by different methods.
The cohomology groups of $T^N \backslash A$ have a natural
structure of a right module over the group ring of the space
symmetry group of the pattern and can be decomposed in a direct
sum of its irreducible representations. An example of such
decomposition is shown for the Ammann-Kramer tiling.
\end{abstract}
\section{Introduction}
One of the distinct features of crystalline structures is that
they are characterized by discrete parameters, in addition to
continuous ones. Examples of such discrete parameters are lattice
symmetry classes, numbers of atoms in the unit cell, occupancies
of Wyckoff positions etc. Taking into account the role of discrete
parameters in our understanding of the structure, it is appealing
to find similar parameters for quasicrystals. Certain of them
could be obtained as a generalization of the discrete parameters
specific for crystals in the framework of the ``cut-and-project''
model. This is the case e.g. for the symmetry class of the
underlying high-dimensional lattice \cite{lattices} or for the
number of atoms in the unit cell (which is replaced by the
homology class of the atomic surface \cite{density}). The efforts
to develop a more systematic approach to the problem have lead to
a promising concept of mutual local derivability (MLD) \cite{MLD}.
Unfortunately, as for now there is little progress in systematic
classification of distinct MLD classes.
\par
An alternative approach to classification of quasicrystals is
based on the notion of the {\em hull} of a quasiperiodic
structure. The concept of hull originated from the works by
Bellissard \cite{bellissard-1} on $C^*\mbox{-algebras}$ of
observables in solid state physics. Bellissard conjectured that
this algebra includes a crossed product of the algebra of
functions on a topological space (called ``the hull'') with the
group of translations acting on it. In many cases, including the
one-particle Schr\"odinger equation in a quasiperiodic potential,
the hull can be described explicitly. The quasiperiodic patterns
of the same MLD class have homeomorphic hulls, which allows one to
characterize quasicrystals by algebraic topological invariants of
their hulls. In this paper we show that some of these invariants,
namely the cohomology ring of the hull may also occur in the study
of the matching rules of quasicrystals.
\par
Before proceeding any further, let us describe briefly the
geometric constructions used in the paper. Following the so-called
``cut-and-project'' method, a quasiperiodic point set is obtained
as an intersection of $d\mbox{-dimensional}$ affine subspace
$E_\parallel \subset \R^N$ with a periodic arrangement of
$(N-d)\mbox{-dimensional}$ manifolds (with boundary) in $\R^N$.
The space $E_\parallel$ is usually referred to as a ``physical
space'', or ``cut'', and the $(N-d)\mbox{-dimensional}$ manifolds
are called ``atomic surfaces''. One can define affine coordinates
on $\R^N$ in such a way that the periodic translations of the
arrangement of atomic surfaces correspond to the vectors with
integer coefficients. The space $\R^N$ can be factored by integer
translations, yielding the $N\mbox{-dimensional}$ torus $T^N$. We
also assume that $E_\parallel$ is not contained in any proper
rational subspace of $\R^N$, hence its image under the natural
projection $\pi: \R^N \rightarrow T^N$ fills densely the torus
$T^N$.
\par
In this paper we consider polyhedral atomic surfaces only. In
order to simplify the proofs we also make several other
non-essential assumptions. In particular, we require that all
connected parts of the atomic surface be flat and parallel to an
$(N-d)\mbox{-dimensional}$ affine subspace $E_\bot \subset \R^N$.
The $\R^N$ is furnished with a Euclidean metric, such that
$E_\parallel$ and $E_\bot$ are perpendicular. When this does not
lead to confusion, we will implicitly switch between $\R^N$ and
$T^N$. In particular, we will use symbols $E_\parallel$ and
$E_\bot$ to designate subspaces in the local coordinate system on
$T^N$. The term ``atomic surface'' will also signify the
submanifold $S \subset T^N$ obtained by the natural projection of
atomic surfaces from $\R^N$. Likewise, we will speak about
translations and convolutions in $T^N$ implying the operations in
the universal cover of $T^N$. The same applies to the definition
of ``piecewise-linear'' (PL) subspaces of $T^N$.
\section{Matching rules and obstacles}
\label{matching} The term ``matching rules'' is usually taken to
mean the set of local constraints on a pattern (a tiling or a
discrete set of points) guaranteeing its global quasiperiodicity.
One can distinguish two approaches to the construction of matching
rules. One approach, which was historically the first, is based on
the scaling symmetry of the quasiperiodic pattern
\cite{penrose,debruijn}. The other one is built upon a more
physical idea of propagation of the quasiperiodic order and leads
to the topological formulation of the matching rules
\cite{katz88,KatzGratias,Katz}. In this section we briefly recall
the derivation of the latter approach.
\par
\begin{figure}\begin{center}
  \epsfbox{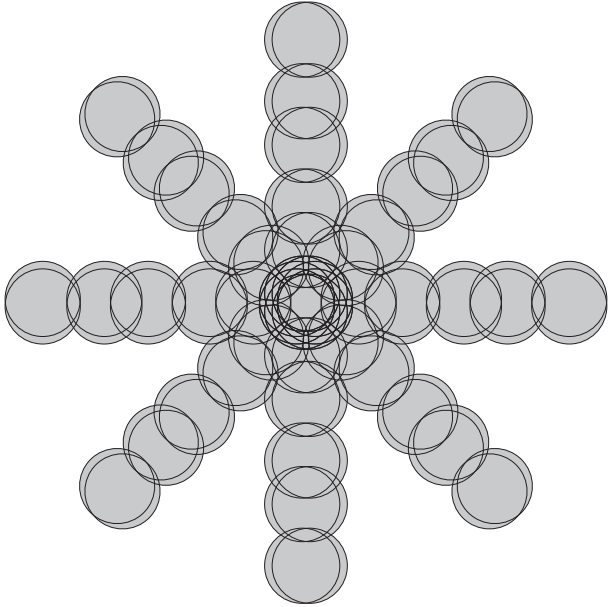}\end{center}
  \caption{\label{stable} The globally connected net formed by
  $R\mbox{-discs}$ centered at the points where a singular cut
  crosses the boundary of the atomic surface of an undecorated
  Ammann octagonal tiling. This cut passes through the vertices
  of the atomic surface.}
\end{figure}

From the very beginning of the study of quasicrystals it has
become obvious that their stability is closely related with the
possibility of propagation of information about the local phason
coordinate. In particular, the stability requires that the places
at which the structure undergoes reconstruction under a uniform
phason shift be arranged in a special way. Namely, when the
magnitude of the phason shift tends to zero, the minimal distance
between the places where the structure is rearranged should not
grow indefinitely, because otherwise no physical mechanism could
guarantee the simultaneousness of the rearrangements
\cite{levitov}. More precisely, there should exist such positive
number $R$, that the union of disks of radius $R$, centered at the
places where the rearrangements occur, form a globally connected
net for any finite uniform phason shift (see Fig. \ref{stable}).
In the general case, the geometry of this net could be quite
complicated. However, we shall restrict our consideration to an
important special class of structures described by flat atomic
surfaces with polygonal boundary. This class includes in
particular the so-called ``model sets'' \cite{Meyer}. In this case
the rearrangements of atoms under a uniform phason shift occur
only when the cut crosses the boundary $\partial S$ of the atomic
surface; this boundary thus plays a crucial role in the
propagation of the quasiperiodic order. In particular, it can be
shown  that the matching rules impose certain constraints on the
orientation of the faces $F_i$ of the boundary \cite{KatzGratias}.
Roughly speaking, the orientation of each face $F_i$ is such that
a {\em singular cut}, crossing it at one point, will cross it at
an infinite number of points. These points form an
$R\mbox{-dense}$ set \cite{r-dense} in a hyperplane in the space
of the cut, as can be seen on Fig. \ref{stable}.
\par
It is important to note, that the rearrangements of the
quasiperiodic pattern under the action of the uniform phason shift
occur simultaneously on the entire net of the Fig. \ref{stable}.
Since such rearrangements do not break the perfect quasiperiodic
order, the matching faults may occur only at the places where the
synchronization of rearrangements is broken at distances smaller
than some finite $R$. The idea of the topological description of
the matching rules stems from an observation that such defects can
be produced if one allows the cut to undulate. In this case, the
matching faults would correspond to intersections of the
undulating cut with the set $Y_R = \partial S + B_R^{\parallel}$,
where $B_R^{\parallel}$ stands for an $R\mbox{-ball}$ in the
parallel direction. The set $Y_R$ is naturally referred to as
``obstacles'' of ``forbidden set'' \cite{katz88,KatzGratias}. The
obstacles $Y_R$ are said to define strong matching rules if any
map of the physical space into $T^N \backslash Y_R$, satisfying
some mild ``sanity conditions'' (e.g. to be everywhere transversal
to the direction of $E_\bot$) is homotopy equivalent to a perfect
cut.
\par
Let us take a closer look at the obstacles $Y_R$ in the case when
they define strong matching rules. In what follows it will be
convenient to associate with each face $F_i$ of the atomic surface
a pair of unit vectors $({\mathbf n}_i, {\mathbf k}_i)$ defined
(up to a sign) as follows. Consider a singular cut crossing the
face $F_i$. This cut intersects $F_i$ at an infinite number of
points belonging to a hyperplane in $E_\parallel$. We define
${\mathbf n}_i \in E_\parallel$ as a normal to this hyperplane and
${\mathbf k}_i \in E_\bot$ as a normal to the face $F_i$.
\par
\begin{figure}\begin{center}
  \epsfxsize=3in
  \epsfbox{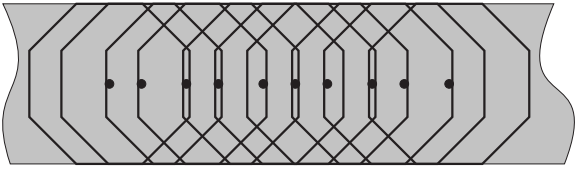}\end{center}
  \caption{\label{thickened_hplane} Thickened hyperplane.}
\end{figure}

The set $Y_R$ can be conveniently represented as
\begin{equation}
\label{obstacle}
 Y_R = \bigcup_{i \in I} t_{R,i},
\end{equation}
where the set $I$ enumerates the faces of the atomic surface and
$T_{R,i}$ is defined as
\begin{equation}
\label{facewise} t_{R,i} = F_i + B_R^\parallel.
\end{equation}
Clearly, for any $R'
> R$, the set of obstacles $Y_{R'}$ defines the same matching
rules. Thus, one can replace the Euclidean norm used to define the
$R\mbox{-balls}$ in the parallel space by any equivalent one. It
is convenient to define the norm of a vector ${\mathbf v} \in
E_\parallel$ as
\begin{equation}
\label{norm} \|{\mathbf v}\| = \max_{i \in I}\left(|{\mathbf n}_i
\cdot {\mathbf v}|\right),
\end{equation}
where the index $i$ enumerates the faces of the atomic surface.
Note that the expression (\ref{norm}) may not define a norm if the
vectors $\{{\mathbf n}_i\}$ span a proper subspace of
$E_\parallel$. If this is the case, we can turn (\ref{norm}) into
a norm by appending to $\{{\mathbf n}_i\}$ the vectors of a basis
of the orthogonal complement to this subspace. The advantage of
the norm (\ref{norm}) over the ordinary Euclidean one is that the
set $Y_r$ defined with the former has an especially simple
geometry. To see this, consider the intersection of a singular cut
with the set $t_{r,i}$ (\ref{facewise}). This intersection is a
union of $r\mbox{-balls}$ with centers belonging to an
$R\mbox{-dense}$ set on a hyperplane perpendicular to ${\mathbf
n}_i$. Note also that an $r\mbox{-ball}$ defined with the norm
(\ref{norm}) is a convex polyhedron and two of its faces are
perpendicular to ${\mathbf n}_i$. As is clear from the figure
\ref{thickened_hplane}, the union of such $r\mbox{-balls}$ for $r$
big enough is a ``thickened'' hyperplane (a set of points
${\mathbf x}\in E_\parallel$ satisfying $a-r \le {\mathbf x}\cdot
{\mathbf n}_i \le a+r$ for some $a$). This can only be possible if
the set $t_{r,i}$ takes the form of a ``thickened torus'':
\begin{equation}
\label{thickened_torus}
 t_{r,i} = T_i + I_i,
\end{equation}
where $T_i$ is an affine subtorus of $T^N$ of codimension two
orthogonal to both ${\mathbf n}_i$ and ${\mathbf k}_i$, $I_i$ is a
segment of length $2r$ parallel to ${\mathbf n}_i$, and the sign
`+' stands for the convolution (see remark in the Introduction).
In what follows we will frequently use the notion of thickened
affine torus, and it is convenient to give it a broader
definition, which will include (\ref{thickened_torus}) as a
special case:
\newtheorem*{def_thickened_torus}{Definition}
\begin{def_thickened_torus}
A thickened affine torus $t$ is a convolution of an affine torus
$T$ with a compact convex subset $B$ of $E_\parallel$:
\begin{equation}
\label{tt}
t=T+B
\end{equation}
\end{def_thickened_torus}
Thus, we have shown that for $r$ big enough, the obstacle $Y_r$ is
a finite union of thickened affine tori (\ref{tt}). Note also that
$Y_r$ can be equipped with a Whitney stratification \cite{whitney}
in such a way that any thickened torus containing a point of a
stratum contains the entire stratum.
\section{Equivalence of cohomology rings of the continuous hull and $T^N \backslash A$}
Following \cite{fhk}, we define the continuous hull $MP$ of the
quasiperiodic pattern as a completion of set of punctured patterns
in the metric of ``approximate match'' $D$ (roughly speaking, two
patterns are separated by the distance $<\epsilon$ in the metric
$D$ if within the ball of radius $1/\epsilon$ the Hausdorff
distance between them is smaller than $\epsilon$; for exact
definition see \cite{fhk}). In this section we establish the
equivalence of the cohomology ring of $MP$ and that of a
complement of $T^N$ to an arrangement of thickened affine subtori
$A$. We start the proof by constructing a sequence of topological
spaces $X_r$ parameterized by a real $r$, and show that $MP$ is
homeomorphic to the inverse limit of this sequence. Then we show
that in the case of quasiperiodic patterns admitting matching
rules, in the homotopy category the limit is attained for a finite
$r_0$. Finally, we demonstrate that the space $X_{r_0}$ is
homotopy equivalent to $T^N \backslash A$. Note that the
representation of the continuous hull of a quasiperiodic pattern
as an inverse limit of topological spaces has already been used in
the literature \cite{anderson_putnam,bel_gam}. Unlike the above
references, the present approach deals directly with the
cut-and-project representation of the quasiperiodic patern, which
allows for a more intuitive description of the limit space. It
should also be mentioned that the role of the matching rules in
convergence of the sequence of cohomology groups of approximating
spaces has been conjectured in \cite{gaehler}.
\par
The set $T^N \backslash Y_r$ represents the origins of the cuts
producing non-singular patterns at least within the
$r\mbox{-disk}$ centered at the origin. In order to include the
singular patterns, one has to add some more points to this space,
which could be done by considering a metric closure of $T^N
\backslash Y_r$. Let us start with the metric on $T^N$ induced by
the Euclidean metric of $R^N$ after factoring it over $\Z^N$ in
the standard position. It induces an inner metric on $T^N
\backslash Y_r$ \cite{burago} (in this metric the distance between
two points equals the infimum of the lengths of the paths in $T^N$
connecting them and avoiding $Y_r$). Denote the completion of $T^N
\backslash Y_{r_n}$ with respect to this metric by $X_{r_n}$.
Consider now an unbounded monotonously increasing sequence $r_n$
and the inclusion maps $\iota_n': T^N \backslash Y_{r_{n+1}}
\rightarrow T^N \backslash Y_{r_{n}}$. Because $\iota_n'$ do not
increase the distance between points, these maps can be extended
to $X_{r_n}$:
\begin{equation}
\label{x_maps} \iota_n: X_{r_{n+1}}\rightarrow X_{r_n}
\end{equation}
One can define the inverse limit of the maps (\ref{x_maps}):
$$
X=\lim_{\leftarrow} X_{r_n}
$$
together with the corresponding projections $\pi_n: X \rightarrow
X_{r_n}$.
\newtheorem*{cor1}{Corollary 1}
\begin{cor1}
The space $X$ is homeomorphic to the {\it continuous hull} $MP$
defined in \cite{fhk}.\end{cor1}
\begin{proof}Recall that $MP$ is defined as completion of the space of
non-singular patterns with respect to the metric of ``approximate
match'' $D$ of \cite{fhk} (two patterns have a distance lesser
than $\epsilon$ if the Hausdorff distance between their patches of
size $1/\epsilon$ does not exceed $\epsilon$). First of all,
remark that there exist continuous maps $\xi_i: MP \rightarrow
X_i$, satisfying $\xi_n=\iota_n \xi_{n+1}$:
\begin{equation}
\label{inclusions} \xymatrix{ \dots \ar[r]^{\iota_{n+1}} &
X_{r_{n+1}} \ar[r]^{\iota_{n}} & X_{r_n}
\ar[r]^{\iota_{n-1}} &  \dots \ar[r]^{\iota_0} & X_{r_0}\\
& MP \ar[u]^{\xi_{n+1}} \ar[ur]^{\xi_{n}} \ar[urrr]^{\xi_{0}} }
\end{equation}
To define the maps $\xi_i$, consider a point $a \in MP$. This
point is a limit of a sequence of patterns obtained by nonsingular
cuts, which is Cauchy in the metric of ``approximate match''. The
origins of these cuts form a sequence of points $x_n \in NS$,
where $NS = \bigcap_i \left(T^N \backslash Y_{r_i}\right)$. In the
metric of $T^N$ the sequence $x_n$ converges to a point $w \in
T^N$ (which may belong to a singular cut!). Consider the cuts with
the origins belonging to the $w+B_\epsilon^\bot$, where
$B_\epsilon^\bot$ is an open PL $\epsilon\mbox{-ball}$ in
$E_\bot$. Those of them, which cross $\partial S$ at the distance
less or equal to $r_k$ from the origin, divide $B_\epsilon^\bot$
in a finite number of open polyhedral pieces $c_j$. There exist
$n_\epsilon$ and $j_0$ such that for $n>n_\epsilon$ all points
$x_n$ belong to $w+c_{j_0} \times B_\epsilon^\parallel$, where
$B_\epsilon^\parallel$ is an $\epsilon\mbox{-ball}$ in
$E_\parallel$. Consider any two points $x_{n_1}$ and $x_{n_2}$ of
the sequence for which $n_1, n_2 > n_\epsilon$. Since $c_{j_0}
\times B_\epsilon^\parallel$ does not intersect $Y_{r_k}$, the
distance between them in the induced inner metric of $T^N
\backslash Y_{r_{k}}$ is bounded by ${\rm const}\cdot \epsilon$.
Therefore, the sequence $x_n$ is Cauchy in the latter metric in
$T^N \backslash Y_{r_{k}}$ and converges to a point in $X_{r_k}$,
which we set as $\xi_k(a)$. The continuity of $\xi_k$ and
commutativity of (\ref{inclusions}) are obvious.
\par
Consider now the continuous map $\zeta: MP \rightarrow X$,
satisfying $\pi_i \zeta = \xi_i$, which exists by virtue of the
universal property of inverse limits. Since $\xi_n$ separates any
two points $a, b \in MP$ for which $D(a, b)>1/r_n$, the map
$\zeta$ is injective. To establish the surjectivity of $\zeta$,
consider a point $x \in X$. For each $k$, its image $\pi_k(x)$ can
be approximated by a sequence of points $x_{k, i} \in NS \subset
T^N \backslash Y_{r_k}$:
$$
\lim_{i\rightarrow\infty} x_{k, i} = \pi_k(x).
$$
The convergence here is defined in the metric of $T^N \backslash
Y_{r_k}$ and without loss of generality can be assumed to be
uniform in $k$. The inclusions $NS \subset T^N\backslash Y_{r_n}
\subset X_{r_n}$ allows one to consider $x_{k, i}$ as a point in
$X_{r_n}$ for any $n$. Then the diagonal sequence $y_i=x_{i, i}
\in NS$ converges in each $X_{r_k}$ to $\pi_k(x)$ (this follows
from the fact that the maps $\iota_n$ of (\ref{x_maps}) do not
increase distance between points). The patterns obtained by cuts
with origins at the points $y_i$ form a Cauchy sequence in the
metric of ``approximate match''. The limit of this sequence is a
point in $MP$ which we set as $\zeta^{-1}(x)$. Therefore, the map
$\zeta$ is a continuous bijection of a compact Hausdorff space
$MP$ \cite{fhk}, and hence a homeomorphism.\end{proof}
\par
The consideration in the section \ref{matching} suggests that the
homotopy type of $T^N \backslash Y_r$ stabilizes with increasing
$r$, and one would expect the same for $X_r$. This is indeed the
case, more precisely, for the polygonal atomic surfaces the
following result holds (the proof is given in the Appendix):
\newtheorem*{cor2}{Corollary 2}
\begin{cor2}
There exists an arrangement $A$ of thickened affine subtori of
$T^N$ and a finite positive $r$, such that for any $r_{n+1} >
r_{n} \ge r$ there is an inclusion $A \subset Y_{r_n}$ and the
following maps are homotopy equivalences:
\begin{enumerate}
\item{The natural inclusion $\mu_n: T^N \backslash Y_{r_n}
\rightarrow X_{r_n}$.} \item{The inclusion of complements $\nu_n:
T^N\backslash Y_{r_n} \rightarrow T^N\backslash A$.} \item{The map
$\iota_n: X_{r_{n+1}} \rightarrow X_{r_n}$ from (\ref{x_maps}).}
\end{enumerate}
\end{cor2}
An immediate consequence of the above Corollary follows  is that
the homomorphisms of cohomology rings induced by (\ref{x_maps})
$$
\iota_n^*: H^*(X_{r_n}) \rightarrow H^*(X_{r_{n+1}})
$$
are isomorphisms for $r_n \ge r$. Thus the cohomology ring of the
space $X$ equals that of $T^N \backslash A$:
\begin{equation}
\label{cohomring} H^*(X) =
\lim_{\rightarrow}\left(H^*(X_{r_n})\right) = H^*(X_r)= H^*(T^N
\backslash A).
\end{equation}
Combining (\ref{cohomring}) with the Corollary 1 we conclude that
the cohomology ring of the continuous hull $MP$ of a quasiperiodic
pattern admitting matching rules is isomorphic to that of a
complement of $T^N$ to a finite arrangement of thickened affine
tori of codimension 2. This implies in particular that the
cohomologies of $MP$ are finitely generated and can be explicitly
calculated as discussed below.
\section{Cohomology of $T^N\backslash A$}
Our goal is to find the cohomology groups of the complement of the
$N\mbox{-dimensional}$ torus to an arrangement of thickened affine
tori $A$. Let us start with the exact cohomological sequence of
pair $(T^N, T^N \backslash A)$
\begin{equation}
\label{longseq} \xymatrix{H^*(T^N, T^N \backslash A)
\ar[r]^{\alpha^*}& H^*(T^N) \ar[d]^{\beta^*}\\
& H^*(T^N \backslash A) \ar[lu]^{d^*}}
\end{equation}
As a Whitney stratified subspace of a torus, $A$ can be surrounded
by an open mapping cylinder neighborhood $\tilde A$ \cite{mather}.
The mapping cylinder determines a deformation retraction of
$\tilde A$ onto $A$ as well as that of $T^N\backslash A$ onto
$T^N\backslash \tilde A$. As $T^N$ is a compact manifold and
$T^N\backslash \tilde A$ is its closed subspace, one has from
Poincar\'e-Alexander-Lefschetz duality \cite{bredon}
\begin{equation}
\label{duality1} H^i(T^N, T^N \backslash \tilde A)=H_{N-i}(\tilde
A)
\end{equation}
giving due to the deformation retraction property
\begin{equation}
\label{duality} H^i(T^N, T^N \backslash A)=H_{N-i}(A).
\end{equation}
The long exact sequence (\ref{longseq}) together with the duality
relation (\ref{duality}) links the cohomologies of $T^N\backslash
A$ with the homologies of $A$. This is not yet sufficient to
relate $H_{N-i-1}(A)$ with $H^i(T^N\backslash A)$ in each
dimension (this would be the case if the homologies of the
surrounding space vanished in several adjacent dimensions, as is
the case for a sphere, yielding Alexander duality). However, if
the rank of the homomorphism $\beta^*$ is known, it is still
possible to separate the dimensions in the sequence
(\ref{longseq}). Indeed, (\ref{longseq}) could be split in
five-term exact sequences:
\begin{equation}
\label{fiveterm} \fl \xymatrix@C=1.5em{0 \ar[r] & {\rm
Im}(\beta^{n-1}) \ar[r] & H^{n-1}(T^N\backslash A) \ar[r]^{d^{n}}
& H_{N-n}(A) \ar[r]^{\alpha^n} & H^n(T^N) \ar[r] & {\rm
Im}(\beta^{n}) \ar[r] & 0},
\end{equation}
yielding the following equation on Betti numbers:
\begin{equation}
\label{betty} b_{n-1}(T^N\backslash A)=b_{N-n}(A)+c_{n-1}+c_n-{N
\choose n},
\end{equation}
where $c_n$ stands for the rank of the map
\begin{equation}
\label{restriction}
 \beta^n: H^n(T^N) \rightarrow H^n(T^N
\backslash A).
\end{equation}
Thus, the ranks of cohomology groups of $T^N \backslash A$ are
determined by that of the homology groups of $A$ and the ranks of
the maps (\ref{restriction}). To obtain the latter remark that by
exactness of (\ref{fiveterm}), the kernel of $\beta^n$ is
isomorphic to the image of $\alpha^n$. On the other hand,
$\alpha^n$ is by Poincar\'e duality equal to the map $H_{N-n}(A)
\rightarrow H_{N-n}(T^N)$ induced by inclusion $A \subset T^N$.
\section{Homology of an arrangement of affine tori}
The space $A$ defined in Corollary 2 is in general case an
arrangement of {\em thickened} affine tori. However, as mentioned
in Appendix, in many cases this space can be collapsed to an
arrangement of ordinary affine tori, which simplifies the
computation significantly. In this section we assume that $A$ is
already collapsed to such an arrangement.
\par
The homology groups of an arrangement of affine tori could be
conveniently computed using the method of simplicial resolutions
(see e.g. \cite{vassiliev}, although we follow here a slightly
modified version of the method). With this technique, instead of
the arrangement $A$, one considers its {\em resolution space}
$A^\Delta$, which has the same homotopy class as $A$. The explicit
construction of $A^\Delta$ is as follows. Let us associate with
the arrangement $A$ a combinatorial object $L(A)$ called an {\em
intersection poset}. The elements of the intersection poset $x \in
L(A)$ correspond to connected components of nonempty intersections
of the tori constituent the arrangement $A$, and the partial order
is given by reverse inclusion. Note that each nonempty
intersection of affine tori is itself a disjoint union of affine
tori (we treat a point as a special case of zero-dimensional
torus). Consider an abstract simplex $\Delta$ with vertices
enumerated by maximal chains of $L(A)$. For each $x \in L(A)$, the
maximal chains containing $x$ define a face of $\Delta$, which we
denote by $\Delta_x$. Let also $t_x \subset T^N$ stand for the
affine torus corresponding to $x$. Then the space of the
simplicial resolution of $A$ is defined as
\begin{equation}
\label{simplicial} A^\Delta=\bigcup_{x \in L(A)} t_x^\Delta,
\end{equation}
where
\begin{equation}
\label{t_delta} t_x^\Delta=t_x \times \Delta_x
\end{equation}
and the corresponding projection $h: A^\Delta \rightarrow A$ is
induced by the projection of $T^N \times \Delta$ onto the first
component.
\par
\newtheorem*{cor3}{Corollary 3}
\begin{cor3}The projection $h: A^\Delta \rightarrow A$ is a homotopy equivalence.
\end{cor3}
\begin{proof}
First of all, let us show that for any point $a \in A$, the
space $h^{-1}(a)$ is contractible. By construction, $h^{-1}(a)$ is
a simplicial set:
\begin{equation}
\label{layer} h^{-1}(a)=a \times \bigcup_{y\in L_a} \Delta_y,
\end{equation}
where $L_a=\{y \in L | a\in t_y\}$. Notice that there exists a
maximal element $x \in L_a$ defined by the condition
$t_x=\bigcap_{y\in L_a} t_y$. Obviously, for any subset $\{y_i\}
\subset L_a$ satisfying $\bigcap_{i} \Delta_{y_i} \neq
\varnothing$ the elements $y_i$ form a chain, which can always be
extended by including $x$. In other words, any nonempty
intersection of simplices $\Delta_y$ in (\ref{layer}) contains at
least one vertex of $\Delta_x$. Consider a vertex $v \in
\bigcup_{y\in L_a} \Delta_y$, which does not belong to $\Delta_x$.
The intersection of all simplices $\Delta_y$ containing $v$ is
nonempty and thus contain at least one vertex $v' \in \Delta_x$
and hence the entire edge $[v v']$. Collapsing $[v v']$ towards
$v'$ defines a deformation retraction of the entire simplicial set
(\ref{layer}) onto its subset obtained by removing the vertex $v$.
This operation can be repeated to eliminate other vertices not
belonging to $\Delta_x$, which proves the contractibility of
(\ref{layer}).
\par
Recall now that $A$ is a Whitney stratified space. By
construction, the set $L_a$ does not depend on the position of the
point $a$ in the stratum. In other words, over each stratum, the
resolution space $A^\Delta$ has a structure of a trivial bundle
with contractible layer. This observation enables us to follow the
proof of Lemma~1 from \cite{vassiliev}, \S III.3.3. Namely,
consider a triangulation of $A$, which exists due to
\cite{goresky}. The interior of each simplex $\sigma$ of
triangulation is contained within a stratum. Hence the space
$h^{-1}(\sigma)$ also has a structure of trivial bundle with a
contractible layer. Then the projection $h$ can be decomposed as
\begin{equation}
\label{decomp} h=h_n \circ \dots \circ h_1 \circ h_0,
\end{equation}
where $h_k$ contracts the layers over the interior points of
$k\mbox{-dimensional}$ simplices of the triangulation ($h_k$ are
continuous because the layers over the boundary of the simplex are
already contracted). The maps $h_k$ from (\ref{decomp}) are
homotopy equivalences, which proves that $h$ is also a homotopy
equivalence.
\end{proof}
\par
At the first glance, the simplicial resolution only replaces an
arrangement of tori by the union (\ref{simplicial}) of a bigger
number of more complex objects (\ref{t_delta}). However, these
objects intersect each other in a more simple way. In particular,
$t_x^\Delta \cap t_y^\Delta$ is nonempty iff $x$ and $y$ are
comparable. In a similar manner, several spaces (\ref{t_delta})
have nonempty intersection iff the corresponding elements of
$L(A)$ form a chain. In this case the intersection has the form
\begin{equation}
\label{intersection}
\bigcap_i t_{y_i}^\Delta = t_{\max(y_i)}
\times \delta,
\end{equation}
where $\delta$ is a face of $\Delta$. Because the comparable
elements in $L(A)$ correspond to tori of different dimensions, the
maximal number of intersecting spaces $t_{y_i}^\Delta$ in
(\ref{intersection}) cannot exceed $N+1$. This limits the number
of non-zero columns in the corresponding Mayer-Vietoris double
complex to $N+1$. Actually this number is even smaller -- it
equals 2 for two-dimensional patterns and 3 for the icosahedral
Ammann-Kramer tiling.
\section{Two-dimensional patterns}
In the case of two-dimensional quasiperiodic patterns admitting
matching rules the space $A$ is an arrangement of two-dimensional
affine subtori of a four-dimensional torus. As we shall see, in
all cases of interest, these tori intersect each other
transversally, that is at a discrete set of points. Let $m$ denote
the number of tori in $A$. We also denote by $n_k$ the number of
points at which $k$ affine tori intersect simultaneously. The
simplicial resolution of $A$ yields $m$ spaces which are homotopy
equivalent to two-dimensional tori and $\sum_k n_k$ simplices. All
intersections between these spaces are pairwise, giving $\sum_k k
n_k$ intersection points. The only non-zero groups in the term
$E^1$ of the homology spectral sequence of the corresponding
Mayer-Vietoris double complex are the followings:
\begin{eqnarray}
\label{spectral-2}
E^1_{0,2}=\Z^m  \\
E^1_{1,0}=\Z^{2m}\nonumber\\
E^1_{0,0}=\Z^{m+\sum_k n_k}\nonumber\\
E^1_{1,0}=\Z^{\sum_k k n_k}.\nonumber
\end{eqnarray}
Since the above spectral sequence has only two non-zero columns,
it collapses at the $E^2\mbox{-term}$. The only nontrivial
differential between the groups (\ref{spectral-2}) is $\delta:
E^1_{1,0} \rightarrow E^1_{1,0}$. The rank of this differential
equals $m + \sum_k n_k - p$, where $p$ stands for the number of
connected components of $A$. This yields the following Betti
numbers of $A$:
\begin{eqnarray}
\label{betti-2}
b_2(A)=m\\
b_1(A)=m+p+\sum_k (k-1)n_k \nonumber \\
b_0(A)=p \nonumber.
\end{eqnarray}
To obtain the Betti numbers of $T^N \backslash A$ one also needs
to know the ranks $c_n$ of the maps $\beta^n$ (\ref{restriction}).
Since $A$ does not contain cells of dimension higher than 2, the
maps $\beta^0$ and $\beta^1$ are injective, giving $c_0=1$ and
$c_1=4$. On the other hand, in all cases considered below, any
$0\mbox{-cycle}$ and $1\mbox{-cycle}$ on $T^4$ can be represented
by a cycle on $A$. Therefore $\alpha^3$ and $\alpha^4$ from
(\ref{fiveterm}) are surjective, yielding $c_3=0$ and $c_4=0$. To
obtain the rank of the remaining map $\beta^2: H^2(T^4)
\rightarrow H^2(T^4 \backslash A)$ observe that since $E^1_{11}=0$
and $E^1_{20}=0$, the group $H_2(A)$ is the direct sum of the
groups $H_2$ of $2\mbox{-dimensional}$ tori constituent $A$. This
allows for explicit computation of the image of $\alpha^2$
(\ref{fiveterm}). In all cases considered below except of
undecorated Ammann-Beenker tiling and undecorated dodecagonal
tiling the rank of $\alpha^2$ equals 4, which corresponds to
$c_2=2$. This result is likely to be valid for any two-dimensional
quasiperiodic pattern admitting strong matching rules, because of
the following argument using de Rham cohomologies. The volume
forms $\omega_\parallel$ and $\omega_\bot$ in $E_\parallel$ and
$E_\bot$ are closed $2\mbox{-forms}$ on $T^4$ spanning a
two-dimensional space in $H^2_{\rm DR}(T^4)$. On the other hand,
one can embed $\R^2$ in $T^4$ in directions of either
$E_\parallel$ or $E_\bot$, without intersecting $A$. This suggests
that $\beta^2(\omega_\parallel) \neq 0$ and
$\beta^2(\omega_\bot)\neq 0$, that is the rank of $\beta_2$ is at
least equal to 2. On the other hand, the rank of $\beta^2$ cannot
be bigger than 2, because this would allow for continuous
variation of the ``slope'' of $E_\parallel$ in $T^N \backslash A$,
which is forbidden by the matching rules. Indeed, the
$n\mbox{-dimensional}$ volume forms in $R^N$ are parameterized by
the points of the Grassmann manifold $g_{N, n}$. Since
$\dim(g_{4,2})=4$, the manifold of volume forms has codimension 2
in $H^2_{\rm DR}(T^4)$. If the dimension of ${\rm Im}(\beta^2)$
equals 3, this space would intersect the above manifold in the
general case along one-dimensional curves, which would make
possible a continuous variation of the slope of $E_\parallel$. One
can cite as an example the undecorated versions of octagonal
Ammann-Beenker and dodecagonal tilings, for which the rank of
$\beta_2$ equals 3, and which do not admit matching rules.
\par
Let us illustrate the technique described above by calculating the
Betti numbers for the Ammann-Beenker octagonal tiling. The
``atomic surface'' of this tiling in its undecorated version has
the shape of a perfect octagon. Eight edges of the octagon give
rise to eight thickened affine tori (\ref{thickened_torus}).
However, the tori corresponding to the opposite edges knit
together as $r$ increases. This results in four thickened tori,
which have a nonempty intersection and thus can be collapsed to
four affine tori $t_i$. They could be specified by the following
vectors spanning the corresponding hyperplanes in the universal
covering space of $T^4$:
$$
\begin{array}{c}
t_1: (e_1, e_2-e_4)\\
t_2: (e_2, e_1-e_3)\\
t_3: (e_3, e_2+e_4)\\
t_4: (e_4, e_3-e_1),
\end{array}
$$
and by the condition that they all pass through the origin. Here
$e_i$ stand for the basis vectors and we assume that the torus
$T^4$ is obtained by factoring $\R^4$ over the lattice $\Z^4$ in
the standard position. The above tori intersect at three points:
$$
\begin{array}{l}
\mbox{at }(0, 0, 0, 0): t_1, t_2, t_3, t_4\\
\mbox{at }(0, 1/2, 0, 1/2): t_1, t_3 \\
\mbox{at }(1/2, 0, 1/2, 0): t_2, t_4,
\end{array}
$$
yielding numbers of intersections $n_2=2$ and $n_4=1$. Finally,
combining (\ref{betti-2}) with (\ref{betty}) and using the values
of $c_i$ found above, we obtain the Betti numbers for $T^N
\backslash A$ given in Table \ref{tab:betti-2}.
\par
\begin{figure}\begin{center}
  \epsfxsize=3in
  \epsfbox{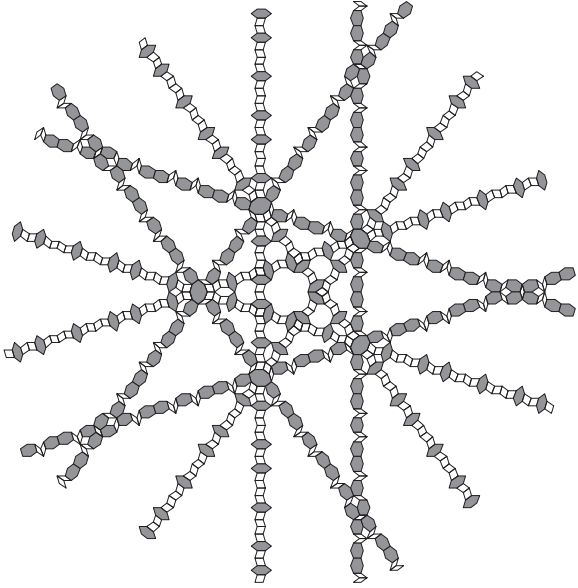}\end{center}
  \caption{\label{p5}Generalized Penrose tiling ($\gamma=5\tau -3$) in a singular position.
  For illustrative purposes only the ambiguously tiled regions (shaded) and the tiles connecting them
  to infinite bands are shown . }
\end{figure}

The computation for other two-dimensional patterns does not differ
qualitatively from the case of Ammann-Beenker tiling. The only
exception is the Penrose tiling, which depends on an extra
parameter $\gamma$ \cite{kleman}. For a generic value of $\gamma$,
the arrangement $A$ consists of 10 affine tori, but when $\gamma
\in \Z[\tau]$ (or, in other words, $\gamma= a + b \tau$), where
$\tau=(5^{1/2}-1)/2$, pairs of parallel thickened tori knit
together. This is illustrated by Figure \ref{p5}, on which a part
of Penrose tiling with $\gamma=5\tau -3$ is shown. Since the
tiling on Figure \ref{p5} is obtained by a singular cut, position
of certain vertices is undefined (the affected tiles are shaded).
The ambiguously tiled regions are aligned along 10 straight lines,
corresponding to 10 thickened affine tori of $Y_r$. However, with
increasing $r$, each pair of parallel lines will form a single
band on the plane of the cut. As a result, the arrangement $A$
consists of only 5 affine tori, all intersecting at the same
point. An infinitesimal variation of $\gamma$ causes displacement
of tori making up $Y_r$ in the direction transversal to the cut,
and they do not knit together anymore. This peculiarity of the
values $\gamma \in \Z[\tau]$ was first observed in \cite{letu}.
Note, however, that we do not see any anomalous behavior of the
cohomology groups for two other classes of $\gamma$, reported in
\cite{GK}, namely $\gamma \in \pm 1/3 + \Z[\tau]$ and $\gamma \in
1/2 + \Z[\tau]$. \fulltable{\label{tab:betti-2}Betti numbers of
$T^N \backslash A$ for various two-dimensional quasiperiodic
patterns. In addition to Betti numbers $b_1$ and $b_2$ the
following parameters of the arrangement $A$ are given: the number
of tori $m$, the number of connected components $p$, the rank
$c_2$ and the numbers of $k\mbox{-wise}$ intersection points
$n_k$. These parameters enter in the formulas (\ref{betti-2}) and
(\ref{betty}).}
\begin{tabular}{lcccccc}
\br Tiling & $b_1$ & $b_2$ & $m$ & $p$ & $c_2$ & Numbers of intersections\\
\mr Ammann-Beenker & 5 & 9 & 4 & 1 & 3 & $n_2=2$, $n_4=1$\\
Ammann-Beenker decorated & 8 & 23 & 8 & 1 & 2 & $n_2=6$, $n_4=1$, $n_8=1$ \\
Penrose ($\gamma \in \Z[\tau]$)& 5 & 8 & 5 & 1 & 2 & $n_5=1$ \\
Penrose ($\gamma$ generic) & 10 & 34 & 10 & 1 & 2 & $n_2=10$, $n_4=5$ \\
dodecagonal & 7 & 28 & 6 & 1 & 3 & $n_2=9$, $n_3=4$, $n_6=1$\\
dodecagonal decorated & 12 & 59 & 12 & 1 & 2 & $n_2=12$, $n_3=8$, $n_4=3$, \\&&&&&&$n_{12}=1$\\
\br
\end{tabular}
\endfulltable
\section{Icosahedral Ammann-Kramer tiling}
The atomic surface of the Ammann-Kramer tiling is the
triacontahedron, obtained as the projection of the unit cube onto
$E_\bot$. Each of 30 faces of the atomic surface gives rise to an
$R\mbox{-dense}$ set of points on a plane in the corresponding
singular cut. The singular cut crossing a face of the
triacontahedron always crosses the opposite face as well. As a
result, the thickened affine tori (\ref{thickened_torus})
corresponding to the opposite faces knit together. Note also that
since a singular cut crossing the triacontahedron at its vertex
also crosses it at all faces, all resulting 15 thickened tori have
a nonempty common intersection. They can also be thinned down to
15 4-dimensional affine tori, as explained at the end of Appendix.
These tori are perpendicular to the two-fold symmetry axes. They
intersect each other at 46 2-dimensional tori, which form four
orbits under the action of the symmetry groups. Two orbits of 15
elements consist of the tori parallel to the two-fold symmetry
axes, one orbit of 10 elements comprise the tori parallel to the
three-fold axes, and the remaining orbit includes 6 tori parallel
to five-fold axes. There are 32 intersection points, forming one
orbit of 30 points and two exceptional points through which pass
all 4-dimensional tori. Since the length of maximal chains of the
intersection poset equals 3, there are only three non-zero columns
in the associated Mayer-Vietoris double complex. The corresponding
homology spectral sequence thus necessarily collapses at the
$E^3\mbox{-term}$. But, as we shall see, the only remaining
non-trivial differential $\partial_2: E^2_{2,0} \rightarrow
E^2_{0,1}$ vanishes because of the symmetry considerations, and
the spectral sequence collapses already at the $E^2\mbox{-term}$
\par
The idea to use the symmetry of the pattern stems from the
observation that there is a naturally defined right action of the
space symmetry group of $T^N \backslash A$ on the cohomologies of
this space. Similarly, one can define a left action of this group
on the homology groups of $A$. This action can be continued onto
the simplicial resolution space $A^\Delta$ and hence on the entire
Mayer-Vietoris double complex. Since the differentials of the
associated homology spectral sequence commute with the action of
the symmetry group, the group action is also defined on all terms
of the spectral sequence. It is natural to decompose the elements
of the spectral sequence in the direct sum of irreducible
representations of the symmetry group (assuming that the
homologies with coefficients in $\R$ are considered). The result
of such decomposition is shown in Table \ref{tab:spectral}. The
symmetry  of the arrangement $A$ is that of the body centered
icosahedral 6-dimensional lattice (note that the symmetry of $A$
is higher than that of the tiling itself). The space group
factored over the translations of the cubic lattice is isomorphic
to $\bi{I}\times\Z_2$. We use the notations of \cite{BG} for the
irreducible representations of $\bi{I}$, while the symmetric and
antisymmetric representation with respect to $\Z_2$ part are
distinguished by adding a prime to the symbol of antisymmetric
representation.
\begin{table}
\caption{\label{tab:spectral}Multiplicities of irreducible
representations of $\bi{I}\times\Z_2$ for the elements of the
spectral sequence $E^2$ for the Ammann-Kramer tiling.}
\begin{tabular}{@{}ccccccccccc} \br
Irrep&Dimension&$E^2_{0,4}$&$E^2_{0,3}$&$E^2_{1,2}$&$E^2_{0,2}$&$E^2_{1,1}$&$E^2_{2,0}$&$E^2_{0,1}$&$E^2_{1,0}$&$E^2_{0,0}$\\
\mr
$A$&1&1&&&1&&1&&&1\\
$A'$&1&&&1&&&1&&&\\
$T_1$&3&&4&&2&3&&1&&\\
$T_1'$&3&&&&&2&2&&&\\
$T_2$&3&&4&&2&3&&1&&\\
$T_2'$&3&&&&&2&2&&&\\
$G$&4&1&4&1&3&4&1&&&\\
$G'$&4&&&1&&2&3&&&\\
$H$&5&2&4&1&3&6&2&&&\\
$H'$&5&&&2&&2&4&&&\\
\br
\end{tabular}
\end{table}
\par
As may be seen from Table \ref{tab:spectral}, no irreducible
representation occurs in both $E^2_{2,0}$ and $E^2_{0,1}$. Hence,
no nontrivial differential map can exist between these groups. As
there are no other potentially nontrivial differentials at $E^2$,
the spectral sequence collapses at the $E^2\mbox{-term}$. The
elements of $E^2$ thus correspond to the summands of the graded
modules associated with the homology groups $H_*(A)$. Since the
inclusion maps of the corresponding filtration of $H_*(A)$ commute
with the action of the symmetry group, the Table
\ref{tab:spectral} also defines the decomposition of $H_*(A)$ into
irreducible representations. Recall, however, that our goal is to
compute the cohomology groups of $T^6\backslash A$, which are
related with $H_*(A)$ by the exact sequence (\ref{fiveterm}). The
symmetry group acts on all elements of (\ref{fiveterm}) (the right
action on the homology groups should be defined as the left action
of the inverse element), and this action commutes with the maps of
(\ref{fiveterm}). Hence, projections of the exact sequence
(\ref{fiveterm}) onto irreducible representations of the symmetry
group can be considered independently. The table \ref{tab:beta}
shows the decomposition of various terms of (\ref{fiveterm}) into
irreducible representations (note that the maps $\beta^k$ are zero
for $k\ge 4$). \fulltable{\label{tab:beta}Multiplicities of
irreducible representations of $\bi{I}\times\Z_2$ for the elements
of the exact sequence (\ref{fiveterm})  for the Ammann-Kramer
tiling.}
\begin{tabular}{cccccccccccc}
\br && \centre{10}{Irrep multiplicities}\\
Module & Dimension& \crule{10} \\
&  & $A$ & $A'$ & $T_1$ & $T_1'$ & $T_2$ & $T_2'$ & $G$ & $G'$ & $H$ & $H'$\\
\mr
${\rm Im}(\beta^0)$ & 1 & 1 &   &   &   &   &   &   &   &   &   \\
${\rm Im}(\beta^1)$ & 6 &   &   & 1 &   & 1 &   &   &   &   &   \\
${\rm Im}(\beta^2)$ & 6 &   &   & 1 &   & 1 &   &   &   &   &   \\
${\rm Im}(\beta^3)$ & 2 & 2 &   &   &   &   &   &   &   &   &   \\
$H^0(T^6)$ & 1 & 1 &   &   &   &   &   &   &   &   &   \\
$H^1(T^6)$ & 6 &   &   & 1 &   & 1 &   &   &   &   &   \\
$H^2(T^6)$ & 15 &   &   & 1 &   & 1 &   & 1 &   & 1 &   \\
$H^3(T^6)$ & 20 & 2 &   &   &   &   &   & 2 &   & 2 &   \\
$H^4(T^6)$ & 15 &   &   & 1 &   & 1 &   & 1 &   & 1 &   \\
$H^5(T^6)$ & 6 &   &   & 1 &   & 1 &   &   &   &   &   \\
$H^6(T^6)$ & 1 & 1 &   &   &   &   &   &   &   &   &   \\
\br
\end{tabular}
\endfulltable
This decomposition together with the data from Table
\ref{tab:spectral} gives the final answer for the cohomology
groups of $T^6 \backslash A$ for the Ammann-Kramer tiling, as
shown in the Table \ref{tab:ico}. \fulltable{\label{tab:ico}Betti
numbers and multiplicities of irreducible representations of
$\bi{I}\times\Z_2$ for cohomology groups of $T^6\backslash A$ for
the Ammann-Kramer tiling.}
\begin{tabular}{cccccccccccc}
\br && \centre{10}{Irrep multiplicities}\\
Cohomology & Betti& \crule{10} \\
group & number & $A$ & $A'$ & $T_1$ & $T_1'$ & $T_2$ & $T_2'$ & $G$ & $G'$ & $H$ & $H'$\\
\mr $H^0(T^6 \backslash A)$ & 1 & 1 &   &   &   &   &   &   &   &   &   \\
$H^1(T^6 \backslash A)$ & 12 & 1 &   & 1 &   & 1 &   &   &   & 1 &   \\
$H^2(T^6 \backslash A)$ & 72 &   & 1 & 5 &   & 5 &   & 3 & 1 & 3 & 2 \\
$H^3(T^6 \backslash A)$ & 181 & 4 & 1 & 4 & 4 & 4 & 4 & 7 & 5 & 10 & 6 \\
\br
\end{tabular}
\endfulltable
The Betti numbers obtained this way differs by one in dimensions 2
and 3 from those reported in \cite{fhk}.
\section{Summary and discussion}
In this paper we have shown that the cohomology ring of the
continuous hull of a quasiperiodic pattern is isomorphic to that
of a complement of a torus to an arrangement of thickened affine
subtory. This fact can be used to compute the cohomology of the
hull. The calculations confirm the previously obtained results in
most cases, with exception of the generalized Penrose tiling and
Ammann-Kramer tiling. The reason for these discrepancies is still
unclear.
\par
It should be emphasized, that the method of this paper could be
applied to other homotopy invariants of the hull as long as they
correspond to continuous functors from the homotopy category. In
particular, the $K\mbox{-theory}$ of the hull should be isomorphic
to the that of $T^N \backslash A$. This is an important
observation since $K\mbox{-groups}$ of the hull are used to label
the gaps in the spectra of quasiperiodic potentials
\cite{bellissard-2,kellendonk}. The isomorphism between
$K\mbox{-groups}$ of the hull and of $T^N \backslash A$ could
provide us with a more intuitive geometric view of the nature of
the gaps and spectral projections.
\par
The cohomologies of $T^N \backslash A$ also provide a way for
classification of topological matching faults in the quasicrystals
\cite{kleman-2}. This can be illustrated by the following example.
Let us consider a large spherical patch of quasicrystal,
containing no matching faults near the surface. The question
arises: is it possible to tell just by looking at the surface that
there are matching faults in the interior of the patch? In some
instances the answer may be positive. Indeed, as the surface layer
is free of matching faults, one can define the map $S^2
\rightarrow T^N \backslash A$, where $S^2$ represents the surface
of the patch. If there are no matching faults in the entire patch,
this map can be continued to the three-dimensional disk. Clearly,
if the homotopy type of the map $S^2 \rightarrow T^N \backslash A$
is nontrivial, such continuation is not possible. Hence, the
elements of $\pi_2(T^N \backslash A)$ correspond to irremovable
point-like matching faults; in the same manner the linear defects
are characterized by the elements of $\pi_1(T^N \backslash A)$.
Therefore, each element of cohomology groups of $T^N \backslash A$
defines an integer-valued function on the matching faults through
the dual of Hurewicz map $H^n(T^N \backslash A) \rightarrow
\hom(\pi_n(T^N \backslash A), \Z)$. These values could be
interpreted as ``topological charges'' of matching faults. \ack
The author is grateful for stimulating discussions to
G.~Abramovici, A.~Katz, J.~Kellendonk, N.~Mnev and P.~Pushkar'.
\section{Appendix}
This appendix contains the proof of the Corollary 2. To begin
with, let us consider the compact space  $Y_r$ as a polyhedron in
a local PL-topology of $T^N$. Then, there exists a regular
neighborhood of $Y_r$ in $T^N$, which we denote by $N_{Y_r}$. The
complement to its interior $T^N \backslash \mathring{N}_{Y_r}$ is
a subspace of $T^N \backslash Y_r$, and could also be considered
as a subspace of $X_r$. Owing to the properties of regular
neighborhoods, one can define a deformation retraction of $\rho:
T^N\backslash Y_r \rightarrow T^N\backslash \mathring{N}_{Y_r}$.
The question arises, whether it is possible to extend $\rho$ on
$X_{r}$ or in other words whether there exists a deformation
retraction $\rho'$ making the following diagram commutative:
\begin{equation}
\label{extend} \xymatrix{X_r \ar@{.>}[dr]^{\rho'} \\
T^N\backslash Y_r \ar[u]^{\mu} \ar[r]^\rho & T^N\backslash
\mathring{N}_{Y_r}}.
\end{equation}
The answer depends on the topology of the embedding of $Y_r$ in
$T^N$, because in general the metric completion modifies the
homotopy type of the complement (e.g. for the complements to
manifolds of codimension bigger than one). The following condition
is sufficient for extension of $\rho'$ on $X_r$:
\newtheorem*{lem}{Lemma}
\begin{lem}
If any point $y \in Y_r$ has a simplicial neighborhood $N_y$ in
$T^N$, such that $N_y \bigcap \left\{T^N \backslash Y_r \right\}$
is collapsible in a {\em finite} number of steps on $\partial N_y
\bigcap \left\{T^N \backslash Y_r \right\}$ then there exists a
deformation retraction $\rho: T^N\backslash Y_r \rightarrow
T^N\backslash \mathring{N}_{Y_r}$ for which the diagram
(\ref{extend}) can be completed by $\rho'$.
\end{lem}
\begin{proof}
Let $(K, L)$ be the triangulations of $(N_{Y_r}, Y_r)$, which
exist by virtue of the simplicial neighborhood theorem
\cite{rourke-sanderson}. By the condition of the Lemma, for each
vertex $a$ of $L$ there exists a collapse
\begin{equation}
\label{collapse} N(a, K) \backslash N(a, L) \searrow \partial N(a,
K) \backslash \partial N(a, L),
\end{equation}
where $N(a, K)$ and $N(a, L)$ stand for simplicial neighborhoods
of $a$ in $K$ and $L$ respectively. The composition of collapses
(\ref{collapse}) for all vertices of $L$ gives a collapse
\begin{equation}
\label{lipschitz} N_{Y_r} \backslash Y_r \searrow \partial
N_{Y_r},
\end{equation}
yielding a deformation retraction $\rho: T^N\backslash Y_r
\rightarrow T^N\backslash \mathring{N}_{Y_r}$. As a composition of
finite number of simplicial maps of finite simplicial complexes,
the collapse (\ref{lipschitz}) satisfies the Lipschitz condition.
Hence, any Cauchy sequence in $T^N\backslash Y_r$ remains Cauchy
during the deformation retraction $\rho$, which allows us to
extend $\rho$ to the metric completion of $T^N\backslash Y_r$.
\end{proof}
\par
The task is now to show that the set $Y_r$ satisfies the condition
of the above Lemma for large enough $r$. According to the remarks
made at the end of the section \ref{matching}, it suffices to
consider the case when $Y_r$ is a union of thickened tori $t_{r,
i}$ (\ref{thickened_torus}). Let us introduce a local coordinate
system on $T^N$ by treating points in a neighborhood of $a \in
T^N$ as vectors ${\mathbf x} \in \R^n$, with $a$ corresponding to
the origin (the space $\R^N$ can be thought of as a universal
covering space of $T^N$). Consider a thickened torus $t_{r, i}$
and let $({\mathbf n}_i, {\mathbf k}_i)$ be the corresponding unit
vectors as defined in the section \ref{matching}. If $a$ is an
interior point of $t_{r, i}$, then the equation of $t_{r, i}$ in
the neighborhood of $a$ is
\begin{equation}
\label{thick_eq} {\mathbf x} \cdot {\mathbf k}_i=0.
\end{equation}
If $a$ lies at the boundary of $t_{r, i}$, then one has to add one
of the following inequalities to the condition (\ref{thick_eq}):
\begin{equation}
\label{thick_ineq} {\mathbf x} \cdot {\mathbf n}_i \ge 0 \;\; {\rm
or} \;\; {\mathbf x} \cdot {\mathbf n}_i \le 0.
\end{equation}
Let now $a$ be an arbitrary point of $Y_r$. It belongs to $t_{r,
i}$ for $i \in I' \subseteq I$, and lies at the boundary of $t_{r,
i}$ for $i \in I'' \subseteq I'$ (the set $I''$ may be empty). One
can choose a neighborhood of $a$ in the form
$B_\epsilon=B^\parallel_\epsilon \times B^\bot_\epsilon$, where
$B^\parallel_\epsilon$ and $B^\bot_\epsilon$ are PL
$\epsilon\mbox{-balls}$ in $E_\parallel$ and $E_\bot$
correspondingly. Our goal is to give an explicit construction of
the collapse $B_\epsilon \bigcap \left\{T^N \backslash Y_r
\right\} \searrow \partial B_\epsilon \bigcap \left\{T^N
\backslash Y_r \right\}$. We begin by cutting $B_\epsilon$ by
hyperplanes $\{{\mathbf x} \cdot {\mathbf k_i}=0 \;|\; i \in I'\}$
and $\{{\mathbf x} \cdot {\mathbf n_i}=0 \;|\; i \in I''\}$. The
resulting cells together with all their faces form a cell complex
$G$, with the underlying space $|G|=B_\epsilon$. It is pertinent
to note that $B_\epsilon \bigcap Y_r$ corresponds to a subcomplex
$H$ of $G$. Furthermore, the complex $G$ is in fact a product of
two cell complexes $G= G^\parallel \times G^\bot$, obtained by
cutting of $B_\epsilon^\parallel$ and $B_\epsilon^\bot$ by the
hyperplanes orthogonal to ${\mathbf n_i}$ and ${\mathbf k_i}$
respectively. For any cell $C \in G^\parallel$ except of maybe
one, which we denote by $C_0$, the space $B_C=C \times
B_\epsilon^\bot$ is cut by one or more of the hyperplanes
(\ref{thick_eq}). Hence, the complement to its intersection with
$Y_r$ is collapsible to the analogous complement of its boundary:
$B_C \backslash \{B_C \bigcap Y_r\} \searrow \partial B_C
\backslash \{\partial B_C \bigcap Y_r\}$. Performing the collapses
in the order of decreasing dimension of cells yields either
$\partial B_\epsilon \backslash \left\{\partial B_\epsilon \bigcap
Y_r \right\}$ if the exceptional cell $C_0$  does not exist or
$(\partial B_\epsilon \bigcup B_{C_0}) \backslash \left\{(\partial
B_\epsilon \bigcup B_{C_0}) \bigcap Y_r \right\}$ otherwise.
Because the interiors of both $B_{C_0} \backslash  \{ B_{C_0}
\bigcap Y_r \}$ and $\partial B_\epsilon \bigcap \left( B_{C_0}
\backslash \{B_{C_0} \bigcap Y_r\}\right)$ are open disks, one
more collapse reduces the latter case to the former, which proves
that the union of thickened tori (\ref{thickened_torus}) satisfies
the condition of the Lemma.
\par
\begin{figure}\begin{center}
  \epsfxsize=3in
  \epsfbox{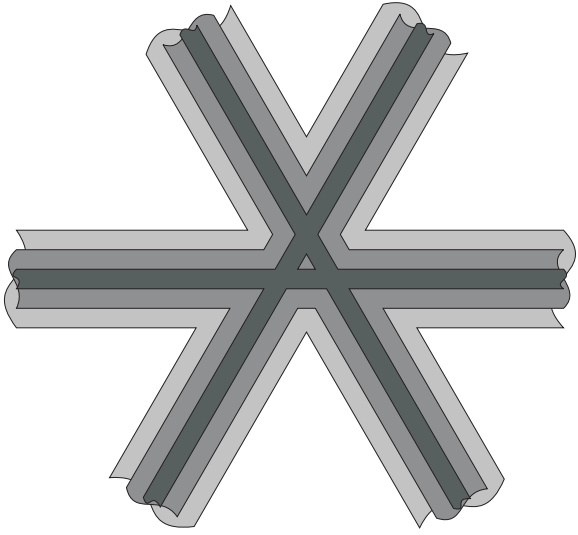}\end{center}
  \caption{\label{stabilization}Intersection of a singular cut with $Y_r$
  for different values of $r$. The shape of the resulting union of thickened
  hyperplanes stabilizes with increasing $r$.}
\end{figure}

It remains to construct an arrangement of thickened affine tori
$A$ in $T^N$ such that $A \subset Y_r$ and that the natural
inclusion $\nu: T^N \backslash Y_r \rightarrow T^N \backslash A$
is a homotopy equivalence. Actually it suffices to show that $Y_r
\searrow A$, because then the regular neighborhood of $Y^r$ in
$T^N$ is also a regular neighborhood of $A$ (see Corollary 3.29
from \cite{rourke-sanderson}). To begin with, consider an
intersection of a singular cut with $Y_r$, which is a finite union
of thickened hyperplanes. As $r$ increases, some faces of the
resulting polyhedron may disappear, but for $r$ big enough the
shape of the polyhedron eventually stabilizes (see Fig
\ref{stabilization}).
Further still, the value of $r$ for which the stabilization occurs
is uniformly bounded by some finite positive $r_0$. This follows
from the observation that the intersection of a singular cut with
$Y_r$ is defined up to translation by the set of faces of
$\partial S$ through which the cut passes and that $\partial S$
has a finite number of faces. Consider now the local structure of
$Y_r$ for $r \ge r_0$. Any point at the boundary of $Y_r$ has a
neighborhood $B_\epsilon$ in which $Y_r$ is locally defined by the
conditions (\ref{thick_eq}) and (\ref{thick_ineq}). The stability
of the shape of the intersection of $Y_r$ with a singular cut
implies that small variations of $r$ correspond to a local
parallel translations of the boundary of $Y_r$. Owing to the
compactness of the boundary of $Y_r$ one can choose a finite
covering of it by neighborhoods $B_\epsilon$ such that the
boundaries of $Y_{r+\delta}$ and $Y_{r-\delta}$ are contained
within it for some $\delta>0$. An appropriate triangulation of
these neighborhoods thus defines a collapse $Y_{r+\delta} \searrow
Y_{r-\delta}$. Hence, for any $r>r_0$ one has $Y_r \searrow
Y_{r_0}$ and the arrangement of thickened tori $A=Y_{r_0}$
satisfies the conditions of the Corollary 2.
\par
The last statement of the Corollary 2 follows from the
commutativity of the following diagram:
\begin{equation}
\label{comm1} \fl \xymatrix{\cdots \ar[r] & X_{r_{n+1}}
\ar[r]^{\iota_n}  & X_{r_n} \ar[r]^{\iota_{n-1}} &
X_{r_{n-1}} \ar[r] & \cdots \\
\cdots \ar[r] & T^N \backslash Y_{r_{n+1}} \ar[r]^{\iota'_n}
\ar[u]^{\mu_{n+1}} \ar[dr]^{\nu_{n+1}} & T^N \backslash Y_{r_{n}}
\ar[r]^{\iota'_{n-1}} \ar[u]^{\mu_{n}} \ar[d]^{\nu_{n}} & T^N
\backslash
Y_{r_{n-1}} \ar[u]^{\mu_{n-1}} \ar[r] \ar[dl]^{\nu_{n-1}} & \cdots \\
& & T^N \backslash A}
\end{equation}
\par
It should be pointed out here that in some cases the thickened
tori constituent the arrangement $Y_{r_0}$ can be ``thinned
down''. In more exact terms, $Y_{r_0}$ can be collapsed to an
arrangement of ordinary affine tori, which may be substituted for
$A$ in Corollary 2. In particular, ``thinning down'' is possible
when all thickened tori in $Y_{r_0}$ have a nonempty intersection.
The quasiperiodic patterns obeying substitution rules also fall in
this category; in this case $A$ may be thought of a result of
``infinite deflation'' applied to $Y_{r_0}$. There are other cases
when $Y_{r_0}$ can be ``thinned down'', including, among others,
the generalized Penrose tiling. It remains unclear, however,
whether this possibility is the common property of all patterns
with polyhedral atomic surfaces.
 \setcounter{section}{1}

\end{document}